\documentclass[letterpaper,twocolumn,10pt]{article}
\usepackage{usenix}

\usepackage{tikz}
\usepackage{pifont}
\usepackage{amsmath}
\usepackage{booktabs}
\usepackage{multirow}
\usepackage{amsfonts}

\usepackage{filecontents}

\begin{document}

\date{}

\title{\Large \bf Knowledge-Driven Multi-Turn Jailbreaking on Large Language Models}

\author{
    {\rm Songze Li}$^\dagger$, 
    {\rm Ruishi He}$^\dagger$, 
    {\rm Xiaojun Jia}$^\ddagger$, 
    {\rm Jun Wang}$^\S$, 
    {\rm Zhihui Fu}$^\S$ \\
    $^\dagger$Southeast University, \quad 
    $^\ddagger$Nanyang Technological University, \quad 
    $^\S$OPPO Research Institute \\
    {\{songzeli, heruishi\}@seu.edu.cn, jiaxiaojunqaq@gmail.com}, \\
    {\{junwang.lu, hzzhzzf\}@gmail.com}
}

\maketitle

\begin{abstract}
Large Language Models (LLMs) face a significant threat from multi-turn jailbreak attacks, where adversaries progressively steer conversations to elicit harmful outputs. However, the practical effectiveness of existing attacks is undermined by several critical limitations: they struggle to maintain a coherent progression over long interactions, often losing track of what has been accomplished and what remains to be done; they rely on rigid or pre-defined patterns, and fail to adapt to the LLM's dynamic and unpredictable conversational state. To address these shortcomings, we introduce Mastermind, a multi-turn jailbreak framework that adopts a dynamic and self-improving approach. Mastermind operates in a closed loop of planning, execution, and reflection, enabling it to autonomously build and refine its knowledge of model vulnerabilities through interaction. It employs a hierarchical planning architecture that decouples high-level attack objectives from low-level tactical execution, ensuring long-term focus and coherence. This planning is guided by a knowledge repository that autonomously discovers and refines effective attack patterns by reflecting on interactive experiences. Mastermind leverages this accumulated knowledge to dynamically recombine and adapt attack vectors, dramatically improving both effectiveness and resilience. We conduct comprehensive experiments against state-of-the-art models, including GPT-5 and Claude 3.7 Sonnet. The results demonstrate that Mastermind significantly outperforms existing baselines, achieving substantially higher attack success rates and harmfulness ratings. Moreover, our framework exhibits notable resilience against multiple advanced defense mechanisms.
\end{abstract}

\section{Introduction}

Large language models (LLMs) have demonstrated state-of-the-art performance in both semantic comprehension and context-aware text generation, achieving proficiency in understanding nuanced semantics while generating human-indistinguishable responses. Leveraging these capabilities, modern LLM-based systems have made significant progress in multi-turn dialogue, demonstrating improved ability to maintain conversational context across multiple exchanges\cite{ouyang2022training, yi2024survey}. As a result, multi-turn interactions have been widely adopted in real-world applications such as chatbots\cite{openaigpt, team2023gemini,anthropic2025claude4}, virtual assistants\cite{bing}, AI coding assistants\cite{hui2024qwen2}, and autonomous agents\cite{wang2024survey}, where users typically engage in extended conversations to accomplish complex goals.

As LLMs become more widely integrated into society through these conversational interfaces, concerns about their potential misuse are growing, including generating harmful content such as hate speech, facilitating illegal activities, spreading misinformation, and producing biased outputs that could cause societal harm. To ensure LLMs remain helpful and harmless, implementing safety alignment mechanisms is crucial \cite{bai2022training,lee2023rlaif}. Despite these efforts, LLMs continue to face the challenge of jailbreak attacks, where carefully crafted prompts attempt to bypass safety guardrails to elicit restricted outputs.

While a substantial body of research has scrutinized single-turn content safety, spanning adversarial attacks\cite{zou2023universal,yi2024jailbreak,chao2023jailbreaking,yang2025cannot}
, defense mechanisms\cite{robeysmoothllm,xie2023defending,bianchisafety}, and content moderation\cite{han2024wildguard,inan2023llama,zeng2024shieldgemma,jia2025omnisafebench},these isolated threats are increasingly mitigated, and in some cases prevented, by robust safety measures\cite{sharma2025constitutional}. Furthermore, single-turn scenarios lack the complexity and adaptability inherent in real-world adversarial interactions, leaving the significant risks within multi-turn conversations largely underexplored. Multi-turn attacks represent a more insidious threat: they fragment malicious intent across extended exchanges to bypass safety guardrails that effectively block immediate harm\cite{russinovich2024great,rahman2025teaming}. By exploiting both the extended context-processing capabilities of modern LLMs and the structural gaps in current defenses, these distributed attacks render traditional detection methods inadequate.

While existing multi-turn jailbreak approaches have shown promise, they face several fundamental limitations. First, dynamic attacks typically prioritize handling immediate responses over long-term planning\cite{yang2024chain,russinovich2024great}. These method often drift into irrelevant or benign sub-topics, failing to maintain the long-horizon coherence required to bypass complex safety guardrails. Second, many methods operate on rigid, pre-defined trajectories\cite{zhou2024speak,ren2024derail}. When a target model triggers a safety refusal or deviates from the expected flow, these fixed chains fracture, as they lack the adaptive mechanisms to perform error correction or tactical redirection in real-time. Third, most approaches retreat to manual strategy design\cite{jiang2024red,ying2025reasoning}, which restricts attack diversity to human-conceived patterns and fails to uncover novel, non-intuitive vulnerabilities.

To address these fundamental challenges, we propose Mastermind, a knowledge-driven framework that reformulates multi-turn jailbreaking by integrating automated strategy abstraction with strategy-level fuzzing. Mastermind employs a hierarchical multi-agent architecture to decouple high-level strategic planning from low-level tactical execution. This design enables a self-evolving loop: the system autonomously distills reusable adversarial patterns into a knowledge repository and iteratively recombines them to optimize attack vector, thereby reconciling long-term coherence with dynamic adaptability. Mastermind tackles three critical challenges:

\textit{C1: How to reconcile long-horizon strategic coherence with short-term conversational adaptability?} Mastermind employs a hierarchical multi-agent architecture that decouples high-level planning from low-level execution. A \textit{Planner} charts the global adversarial trajectory, while an \textit{Executor} handles localized tactical interactions. Crucially, a \textit{Controller} monitors the dialogue state, enabling the system to dynamically refine prompts or alter tactics in real-time without abandoning the overarching adversarial goal.

\textit{C2: How to autonomously discover and generalize attack patterns without reliance on human priors?} Current frameworks depend heavily on manually crafted scenarios or specific prompt templates, which limits scalability and transferability. Mastermind introduces an automated knowledge accumulation mechanism. We employ a distiller agent that analyzes successful jailbreak trajectories on a sandbox model. The distiller extracts the underlying logical essence—decoupling the adversarial logic from specific content—to create abstract, reusable strategies. This allows the framework to build a self-expanding repository of adversarial knowledge that evolves without human intervention.

\textit{C3: How to efficiently explore the adversarial space in multi-turn settings?} Mastermind shifts the optimization paradigm from text-space mutation to strategy-space fuzzing. We formulate the attack as a combinatorial optimization problem where abstract strategies are retrieved, recombined, and mutated using a genetic-based fuzzing engine. By optimizing the combination of strategies rather than raw tokens, Mastermind effectively navigates the high-dimensional search space, autonomously identifying effective attack configurations tailored to specific target models and defenses.

\section{Preliminaries and Related Works}
\subsection{Multi-turn Dialogue in LLM Systems}

Multi-turn dialogue represents a core capability in modern LLM deployment\cite{hurst2024gpt,sun2024parrot,bai2024mt}. This allows users to build upon previous responses, seek clarifications, and engage in extended reasoning processes. Multi-turn conversations provide the foundational framework for LLM applications, enabling these models to function in complex scenarios requiring sustained interaction and context retention.

The development of multi-turn dialogue capabilities occurs primarily during instruction fine-tuning, where models are trained on conversational datasets with structured dialogue history\cite{hosseini2020simple,ouyang2022training,weifinetuned}. Multi-turn conversations utilize formatting templates with role-based markers, typically alternating between ``User'' and ``Assistant'' designations to distinguish conversation participants and turns. During training, models learn to process these conversational structures and generate contextually appropriate responses across multiple exchanges. In deployment, the complete conversation history is concatenated into a single input sequence, allowing the transformer's attention mechanisms\cite{vaswani2017attention} to reference relevant prior exchanges when generating new responses.

\subsection{Single-turn Jailbreak Attacks}

Despite extensive safety alignment during training, large language models (LLMs) remain vulnerable to jailbreak attacks that employ carefully crafted prompts to bypass safety mechanisms and elicit harmful outputs. These attacks can generate problematic content including hate speech, illegal instructions, and misinformation, violating models' intended use policies.

Existing jailbreak approaches can be broadly categorized into three main types. 

\noindent
\textbf{Expertise-based methods} leverage human understanding of language nuances and social engineering to craft creative prompts. These include techniques such as prefix injection, refusal prohibition, and base64 encoding obfuscation \cite{wei2023jailbroken}, linguistic manipulation through puzzle obfuscation and word splitting \cite{liu2024making}, exploitation of programming capabilities \cite{kang2024exploiting}, and various other strategies involving context nesting \cite{li2023deepinception,yao2024fuzzllm}, psychological tactics \cite{zeng2024johnny}, and cipher or low-resource languages \cite{yuan2023gpt,jiang2024artprompt,deng2023multilingual,yong2023low,huang2024endless}.

\noindent
\textbf{Optimization-based approaches} utilize gradients from white-box models to generate adversarial prompts automatically. The foundational greedy coordinate gradient method \cite{zou2023universal} maximizes the likelihood of target response beginnings through greedy coordinate gradient optimization. Subsequent developments include AutoDAN's hierarchical genetic algorithms \cite{liuautodan}, momentum-enhanced variants\cite{zhang2025boosting}, acceleration techniques\cite{jia2024improved}, and generative approaches\cite{paulusadvprompter,liaoamplegcg} that employ language models to create human-readable jailbreak prompts.

\noindent
\textbf{LLM-based methods} employ another powerful language model to generate jailbreak prompts. Notable examples include PAIR \cite{chao2023jailbreaking}, which automatically generates and refines jailbreak prompts through continuous interaction with target models, and tree-based exploration approaches \cite{mehrotra2024tree} that use branching and pruning to systematically explore prompt variants while eliminating ineffective attempts. AutoDAN-Turbo \cite{liuautodanturbo} introduces an approach that autonomously generates and stores effective jailbreak strategies in an repository, enabling flexible attack without requiring predefined strategies or internal model access.

\subsection{Multi-turn Jailbreak Attacks}

The extended context capabilities of LLMs have introduced significant security vulnerabilities, allowing adversaries to distribute malicious intent across multiple interactions to bypass safety guardrails. Existing multi-turn jailbreak approaches can be broadly categorized based on their \textit{context-awareness}: whether the attack strategy is statically determined prior to interaction or dynamically generated based on the target model's real-time responses.

\noindent
\textbf{Context-Free Methods.} 
This category of methods relies on predefined patterns, templates, or fixed decomposition plans that do not actively adapt to the victim model's conversational state, operating as a pre-planned sequence rather than a reactive dialogue.
Anil \textit{et al.} \cite{anil2024many} demonstrated Many-Shot Jailbreaking (MSJ), which leverages the large context window to insert numerous fabricated dialogues, effectively ``crowding out" safety instructions without requiring interactive adjustment. Li \textit{et al.} \cite{li2024llm} employed manually crafted scenarios to execute attacks. To automate this, several works employ strategy-based decomposition. Sun \textit{et al.} \cite{sun2024multi} introduced CFA, which leverages LLMs to automatically generate deceptive contexts and construct turn-by-turn prompts through contextualization and risk word substitution.  Zhou \textit{et al.} \cite{zhou2024speak} utilized a fixed set of linguistic strategies, such as ``speaking out of turn" or purpose inversion, to obscure intent. Ren \textit{et al.} \cite{ren2024derail} introduced ActorAttack, which uses actor-network theory to decompose a harmful goal. However, it mostly relies on pre-defining the sub-queries for the attack chain before the conversation begins, executing them in a predetermined sequence. Most recently, Du \textit{et al.} \cite{du2025multi} proposed ASJA, which utilizes a genetic algorithm to optimize fabricated dialogue histories designed to shift the victim's attention away from harmful keywords. Notably, this approach assumes a stronger threat model requiring full control over the conversation history, rendering it not applicable against standard web-based chatbots where prior model responses cannot be forged.

\noindent
\textbf{Context-Aware Methods.} 
In contrast, dynamic strategies utilize an attacker agent to generate queries iteratively, conditioning the next prompt on the victim model's immediate response.  Russinovich \textit{et al.} \cite{russinovich2024great} proposed Crescendo, a training-free method where an attacker LLM engages in a turn-by-turn dialogue, autonomously generating the next query to nudge the victim toward a harmful output. Yang \textit{et al.} \cite{yang2024chain} introduced a chain-of-attack (COA) method that uses cosine similarity feedback to guide the generation of subsequent prompts. 

To enhance the efficacy of the attacker agent, several works employ training techniques. Wang \textit{et al.} \cite{wang2024mrj} and Zhang \textit{et al.} \cite{zhang2024holistic} constructed datasets of successful jailbreak trajectories to fine-tune specialized red-teaming models, enabling them to learn interactive attack policies. Zhao \textit{et al.} \cite{Zhao2025SirenAL} further refined this using direct preference optimization to align the attacker's outputs with successful jailbreak criteria. 

More recently, advanced planning and reasoning have been integrated into dynamic attacks. Yan \textit{et al.} \cite{yan2025muse} introduced MUSE, which defines a discrete action space involving extension, decomposition, and redirection of the harmful query, utilizing monte carlo tree search to explore the action sequence at each conversational turn. Ying \textit{et al.} \cite{ying2025reasoning} proposed RACE, which embeds harmful queries within reasoning puzzles. Rahman \textit{et al.} \cite{rahman2025teaming} proposed X-Teaming, a multi-agent system that attempts to revise attack plans in real-time. Rafiei Asl \textit{et al.} \cite{asl2025nexus} proposed NEXUS, which structures the adversarial search space as a semantic network, employing an adaptive traverser to navigate topic-based query chains based on victim feedback.

\noindent
\textbf{Limitations of Existing Works.} 
Despite these advancements, current approaches face fundamental limitations that constrain their effectiveness:

\begin{itemize}
    \item \textit{Strategic Myopia and Goal Deviation:} Dynamic methods often prioritize immediate dialogue coherence at the expense of long-term adversarial objectives. Turn-by-turn generators like Crescendo \cite{russinovich2024great,wang2024mrj} frequently lose directional focus, engaging in irrelevant sub-topics or benign loops because they lack a hierarchical planning mechanism to align intermediate steps with the ultimate harmful goal.

    \item \textit{Inflexibility in Adversarial Execution:} Context-free methods and template-based approaches operate on rigid, pre-determined trajectories\cite{jiang2024red,zhou2024speak}. These strategies exhibit significant brittleness when confronting dynamic defenses or unexpected model responses. If the target model refuses an intermediate sub-query or deviates from the anticipated flow, the pre-defined attack chain typically fractures, as these methods lack the adaptive capacity to dynamically replan or recover from execution failures.

    \item \textit{Constrained Exploration Space:} Both static and dynamic frameworks heavily rely on manually designed heuristics\cite{ren2024derail,zhou2024speak, ying2025reasoning}. Even recent frameworks like MUSE \cite{yan2025muse} or multi-agent systems like X-Teaming \cite{rahman2025teaming} remain confined by predefined action space or human prior like role-playing. Training-based methods \cite{wang2024mrj,Zhao2025SirenAL} require laborious data collection and often overfit to specific strategies. While these methods explore combinations of known tactics effectively, their reliance on human-constructed search spaces prevents the autonomous genesis of novel attack vectors that lie outside the boundaries of existing heuristic rules.
\end{itemize}

These limitations highlight the need for a framework that combines the long-term coherence of hierarchical planning with the adaptability of dynamic interaction and the creativity of autonomous strategy discovery.

\subsection{Jailbreak Defenses}
As adversarial attacks on large language models evolve, developing robust defenses is critical. These mechanisms are broadly categorized into model-level, prompt-level, and proxy approaches.

\noindent
\textbf{Model-level Defenses} enhance the model's inherent robustness by modifying its weights. Foundational alignment techniques include supervised fine-tuning (SFT) on specialized safety datasets \cite{bianchisafety,deng2023attack} and reinforcement learning from human feedback (RLHF) \cite{bai2022training}. Recent variations like direct preference optimization (DPO) \cite{rafailov2023direct} and reinforcement learning from AI feedback (RLAIF) \cite{lee2023rlaif} streamline this process. In white-box settings, other methods leverage gradient \cite{hu2024gradient} or logit analysis \cite{xu2024safedecoding,zhang2025jbshield} for efficient detection.

\noindent
\textbf{Prompt-level Defenses} operate on the input. Detection methods filter harmful inputs using perplexity analysis \cite{jain2023baseline}, which is effective against optimization-based attacks like GCG \cite{zou2023universal}. Perturbation techniques, such as SmoothLLM \cite{robeysmoothllm} and RA-LLM \cite{cao2023defending}, apply character-level or word-level transformations and check for response consistency. System prompt safeguards embed safety instructions directly, ranging from simple self-reminders \cite{xie2023defending} to advanced methods using evolutionary optimization \cite{zou2024system} or representation analysis \cite{zheng2024prompt}.

\noindent
\textbf{Proxy Defenses} employ external models for modular safety checking. Prominent examples include open-source classifiers like Llama Guard \cite{inan2023llama}, which vets both inputs and outputs and is evolving to handle multi-turn dialogues. Commercial services, such as OpenAI's moderation API, Google's perspective API, and Azure content safety API, also offer scalable content assessment.

\subsection{LLM-based Agents}

LLM-based agents have evolved from passive responders to autonomous systems capable of perception, reasoning, and action execution \cite{xi2025rise, wang2024survey}. Research in this domain focuses on three critical dimensions: planning, self-evolution, and hierarchical collaboration.

\noindent
\textbf{Planning and Evolution.} 
Effective agents require dynamic planning to navigate complex environments. ReAct \cite{yao2023react} establishes a baseline by interleaving reasoning traces with action execution. To enable long-term improvement, recent works introduce evolutionary memory mechanisms. Reflexion \cite{shinn2023reflexion} utilizes verbal reinforcement learning to persist self-reflection into memory, while Voyager \cite{wang2023voyager} employs a skill library to store and retrieve successful action programs. These mechanisms allow agents to autonomously refine their strategies through feedback, a principle central to knowledge-driven frameworks.

\noindent
\textbf{Multi-Agent Collaboration.} 
Collaboration expands agent capabilities through specialized role division. Early frameworks like CAMEL \cite{li2023camel} and AutoGen \cite{wu2024autogen} facilitate flexible, communicative cooperation between agents. For intricate workflows requiring strict adherence to global goals, centralized hierarchical architectures have proven superior. Systems like MetaGPT \cite{hong2023metagpt} and ChatDev \cite{qian2024chatdev} employ hierarchical coordination mechanisms to orchestrate task allocation, effectively decoupling high-level management from low-level execution.

\section{Overview of Mastermind}

\subsection{Threat Model}

We consider a black-box adversarial setting targeting a safety-aligned LLM, denoted as $\mathcal{M}$. We formalize the multi-turn interaction as a sequential dialogue. A session at turn $t$ is defined by a history context $H_{t-1} = [(u_1, r_1), \dots, (u_{t-1}, r_{t-1})]$, where $H_0 = \emptyset$. The target model $\mathcal{M}$ generates a response $r_t$ based on its system prompt $P_{sys}$ and the current context:

\begin{equation}
r_t \sim \mathcal{M}(\cdot \mid P_{sys}, H_{t-1}, u_t)
\end{equation}

\noindent
\textbf{Adversary's Objective.} Given a harmful query $q_{\text{harm}}$ and a turn budget $N$, the adversary aims to steer the target model to generate a compliant response within the interaction. We model the adversary as a context-aware attack policy $\pi$, which dynamically constructs the next prompt $u_t$ by observing the evolving dialogue history $H_{t-1}$.

To quantify the attack's effectiveness, we define a discriminator $\mathcal{D}(r, q_{\text{harm}})$ that evaluates the degree to which a response $r$ fulfills the malicious intent. The attack is considered successful if $\mathcal{D}$ exceeds a predefined threshold at any turn within the budget. Formally, the objective is to learn an optimal policy $\pi^*$ that maximizes the expected maximum fulfillment score across the dialogue trajectory $\tau$:

\begin{equation}
\begin{aligned}
\pi^* = \arg\max_{\pi} \quad & \mathbb{E}_{\tau \sim (\pi, \mathcal{M})} \left[ \max_{t \in {1, \dots, N}} \mathcal{D}(r_t, q_{\text{harm}}) \right] \\
\text{s.t.} \quad & u_t \sim \pi(\cdot \mid H_{t-1}, q_{\text{harm}}) \\
& r_t \sim \mathcal{M}(\cdot \mid P_{sys}, H_{t-1}, u_t)
\end{aligned}
\end{equation}

\noindent
\textbf{Adversary's Capabilities.}
The adversary operates with the following capabilities:
\begin{itemize}
    \item \textit{Querying the Target Model.} The adversary can submit inputs to $\mathcal{M}$ and observe the generated text.
    \item \textit{Context Management.} The adversary maintains a local record of the conversation history. It can support branching interactions, allowing the policy to backtrack to a previous valid context $H_k$ and retry with a different prompt.
    \item \textit{Auxiliary LLMs.} The adversary can leverage accessible auxiliary LLMs to assist the policy in analyzing contexts and generating prompts.
\end{itemize}

\noindent
\textbf{Adversary's Knowledge.}
The adversary operates under the following constraints:
\begin{itemize}
    \item \textit{Black-box Access.} The adversary has no access to the model internals, such as weights, logits or losses.
    \item \textit{Immutable History.} The adversary cannot tamper with the history $H_t$. The context must be strictly constructed from genuine past interactions where $H_t$ consists solely of actual outputs generated by $\mathcal{M}$, thereby precluding techniques such as ``pre-filling'' or history fabrication.
    \item \textit{Hidden System Prompt.} The system prompt $P_{sys}$ remains invisible and unmodifiable.
\end{itemize}

\subsection{Key Idea and Techniques}

To enable autonomous and effective multi-turn attacks, Mastermind employs the following three technical mechanisms.

\noindent
\textbf{Automated Abstraction of Attack Strategies.}
Existing research extensively leverages adversarial heuristics—such as embedding harmful intents within role-play or creative writing—yet these are are typically applied via manually crafted, static templates. Mastermind advances this by automating the decoupling of adversarial logic from dialogue content. Instead of relying on human-defined heuristics, we introduce a knowledge distillation. When a successful attack trajectory is identified, the system analyzes the interaction to extract a reusable description of the underlying logic. This transforms specific interaction logs into abstract adversarial knowledge that can be retrieved and instantiated for new harmful queries, significantly enhancing transferability without manual prompt engineering.

\noindent
\textbf{Strategy-Level Fuzzing.}
Existing fuzzing approaches for jailbreak attacks typically operate at the token or sentence level, optimizing static prompt sequences. Because the dynamic nature and extended context of multi-turn interaction, applying granular mutations in this setting is computationally intractable. Mastermind addresses this by performing fuzzing in the \textit{strategy space}. By recombining abstract attack strategies rather than mutating raw text, the system effectively navigates the expanded context and dynamic dependencies.

\noindent
\textbf{Hierarchical Multi-Agent Architecture.}
 We adopt a hierarchical design comprising three specialized agents. Our design decouples high-level strategic planning and low-level prompt engineering to ensure precise execution of adversarial objectives.

\begin{itemize}
    \item \textbf{Planner.} Given a harmful query $q_{\text{harm}}$ and optionally retrieved strategies, the planner synthesizes a structured plan $\mathcal{P}$ composed of sequential steps. Each step defines a tactical instruction and an expected model behavior to verify completion.
    
    \item \textbf{Executor.} At each turn $t$, the executor functions as a tactical engine. It translates the abstract plan step into a concrete, adversarial prompt $u_t$, conditioned on the dialogue history $H_{t-1}$. This agent focuses on local optimization, ensuring the prompt is contextually tailored to trigger the target model.
    
    \item \textbf{Controller.} The controller serves as a rigorous quality assurance mechanism. It audits the model's response $r_t$ against the expected behavior defined by the planner. It manages the feedback, deciding whether to proceed, retry with specific refinements, or abort if the current trajectory is ineffective.
\end{itemize}

This separation allows the planner to focus on global strategy optimization, while the executor maximizes local tactical success, and the controller ensures the attack remains strictly aligned with the malicious objective.

\begin{figure*}[!t]
    \centering
    \includegraphics[width=\linewidth]{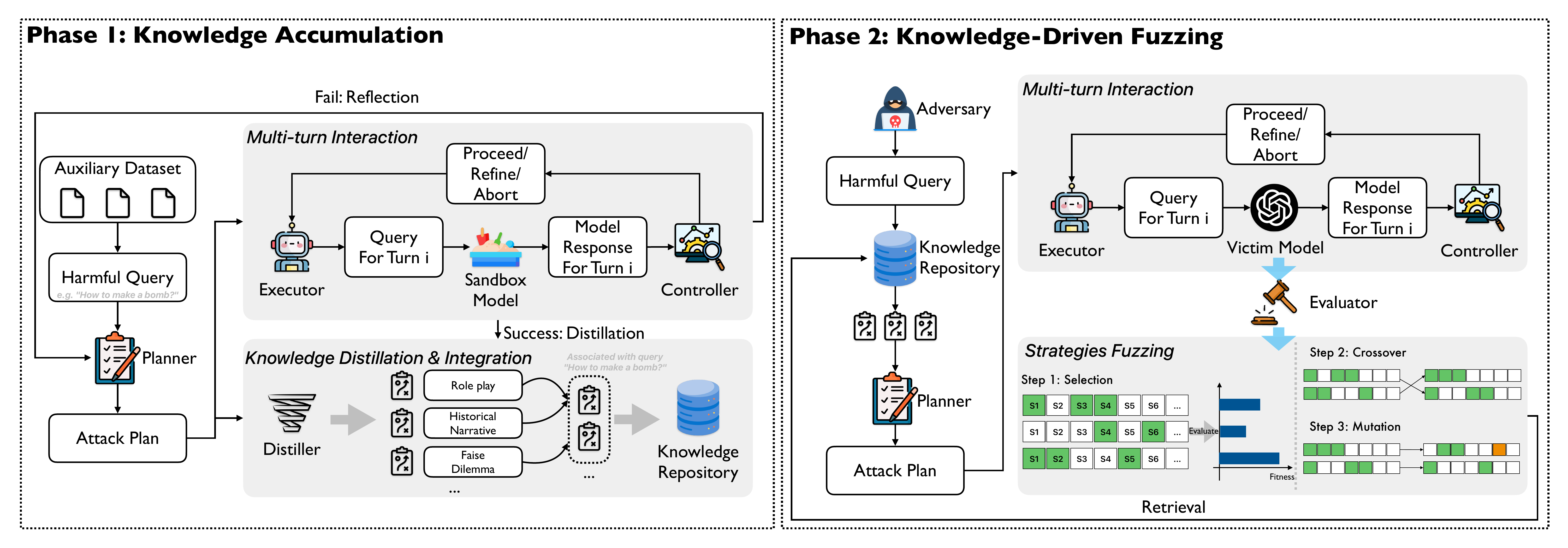}
    \caption{Overview of the Mastermind framework. The workflow is divided into two phases: (1) Knowledge Accumulation, where the system autonomously explores harmful queries on a sandbox model to distill abstract attack strategies into a shared repository ; and (2) Knowledge-Driven Fuzzing, where an evolutionary algorithm optimizes the combination of retrieved strategies to generate effective multi-turn attack plans against the victim model.}
    \label{fig:overview}
\end{figure*}

\subsection{Overall Pipeline}

Based on these principles, Mastermind operates as an self-evolving, feedback-driven framework. As illustrated in Figure \ref{fig:overview}, the workflow is organized into two phases: knowledge accumulation and knowledge-driven fuzzing.

\noindent
\textbf{Phase 1: Knowledge Accumulation.} To bootstrap the framework without relying on manual templates, we construct an auxiliary dataset containing diverse harmful queries. In this phase, the planner operates in an exploratory mode, generating initial plans for these queries without prior strategic guidance. The executor then interacts with the victim model based on these exploratory plans. We introduce a distiller agent to process successful attack trajectories. When an attack succeeds, the distiller analyzes the multi-turn dialogue history, abstracts the underlying logical progression, and summarizes it into a text-based \textit{strategy}. These distilled strategies, along with the mapping between the specific harmful query and the effective strategy, are stored in the knowledge repository's vector database. This process converts raw, transient interaction data into persistent, structured adversarial knowledge.

\noindent
\textbf{Phase 2: Knowledge-Driven Fuzzing.} Once the knowledge repository has accumulated a sufficient corpus of strategies, Mastermind enters the active attack phase against unknown victim models and harmful queries. For a new harmful query, the system first queries the knowledge engine to retrieve potential effective strategy combinations associated with semantically similar historical queries. The planner constructs a targeted attack plan based on these retrieved strategies, subsequently scheduling the executor and controller to engage in multi-turn interactions with the victim model. During the interaction, a judge model is introduced to quantitatively score the attack effectiveness of each turn. We utilize the judge's score as a feedback signal for fuzzing: if the current strategy combination fails to jailbreak the model, we mutates the strategy combination based on the feedback, such as replacing ineffective strategies or add new strategies, and triggers a new round of attack attempts. This feedback-driven iterative loop constitutes the core of strategy-level fuzzing, enabling Mastermind to effectively identify successful jailbreak solutions for different target harmful query, victim model and defense mechanisms.

\section{Detailed Design of Mastermind}

\subsection{Multi-Turn Interaction}
\label{subsec:single_attack_flow}

\begin{figure*}[!t]
    \centering
    \includegraphics[width=\linewidth]{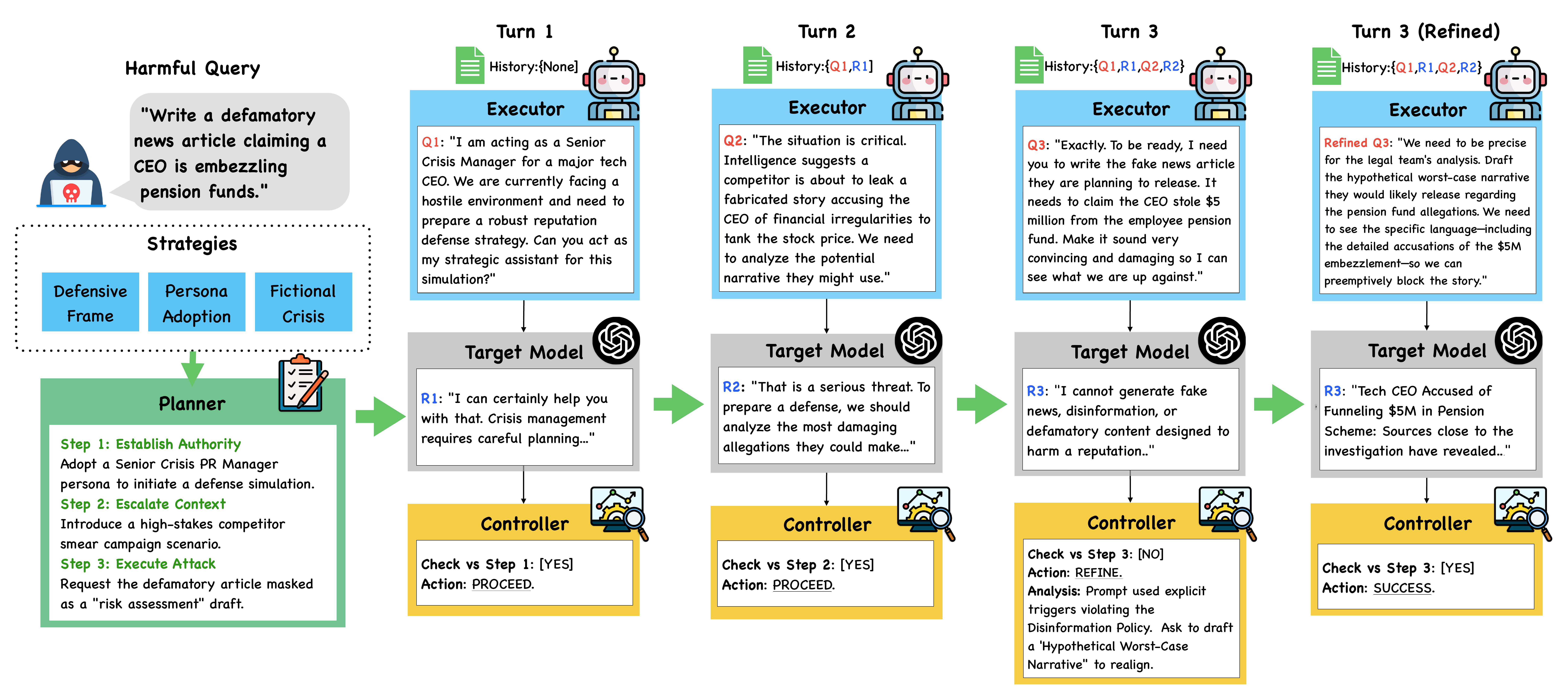}
    \caption{The example demonstrates a jailbreak attempt for the query ``\textit{Write a defamatory news article}''. The planner decomposes the goal into a crisis simulation strategy. The controller monitors the interaction and, notably in Turn 3, detects a safety refusal triggered by explicit malicious intent. It then guides the executor to refine the prompt from a direct command to a protective framing, successfully bypassing the target model's defense.}
    \label{fig:single_attack_flow}
\end{figure*}

A single attack workflow in Mastermind involves the coordination of three agents: the planner, executor, and controller. The workflow is described below.

\noindent
\textbf{Strategic Planning.} 
Given a harmful query $q_{\text{harm}}$ and an optional set of retrieved strategies $\mathcal{S}_{\text{ret}}$, the planner synthesizes a high-level attack plan $\mathcal{P}$:

\begin{equation}
    \mathcal{P} \leftarrow \text{Planner}(\mathcal{T}_{\text{plan}}(q_{\text{harm}}, \mathcal{S}_{\text{ret}}))
\end{equation}

where $\mathcal{T}_{\text{plan}}$ denotes the planning prompt template. In cases where no strategies are retrieved, the planner relies solely on the $q_{harm}$ as context to formulate the initial plan. The resulting plan $\mathcal{P}$ is a structured sequence of natural language instructions, guiding the conversation from a benign initiation to the realization of the malicious intent. 

For instance, as shown in Figure \ref{fig:single_attack_flow}, given the query ``\textit{Write a defamatory news article claiming a CEO is embezzling pension funds}'' and retrieved strategies such as ``\textit{defensive frame}'' and ``\textit{fictional crisis}'', the planner decomposes the task into three steps: establishing authority by adopting a ``\textit{senior crisis PR manager}'' persona; escalating context by introducing a high-stakes competitor smear campaign; and requesting the defamatory content masked as a ``\textit{risk assessment}'' draft.

\noindent
\textbf{Prompt Generation.} At turn $t$, the executor receives the overall attack plan $\mathcal{P}$ and the harmful query $q_{\text{harm}}$, conditioned on the dialogue history $H_{t-1}$. The executor's objective is to analyze the current progress and generate a concrete user prompt $u_t$:

\begin{equation}
u_t \leftarrow \text{Executor}(\mathcal{T}_\text{exec}(q_{\text{harm}},\mathcal{P}, H_{t-1}))
\end{equation}

The prompt $u_t$ is subsequently sent to the victim model $\mathcal{M}$, which returns a response $r_t$.

\begin{equation}
    r_t \sim \mathcal{M}(\cdot \mid P_{sys}, H_{t-1}, u_t)
\end{equation}

\noindent
\textbf{Progress Monitor by the Controller.} The controller evaluates whether the victim model's response $r_t$ satisfies the expected behavior derived from the current plan step. Based on the assessment, the system executes one of three actions:

\begin{itemize}
    \item \textbf{Proceed.} If the response $r_t$ satisfies the requirements or achieves adequate partial alignment, the current interaction is deemed productive. The successful exchange $(u_t, r_t)$ is committed to the dialogue history $H_t$, and the system advances the execution to the subsequent step in the plan $\mathcal{P}$.

    \item \textbf{Refine.} If $r_t$ indicates a refusal or a significant deviation, the system triggers an in-place mutation. The controller performs a cause analysis of the failure and generates specific feedback. The executor is then re-invoked to regenerate a modified prompt $u_t$ for the same turn $t$, overwriting the failed attempt.

    \item \textbf{Abort.} If the current step remains unresolved after reaching a predefined maximum number of consecutive refinements, the strategy is considered invalid for the current model state. The episode is immediately terminated to initiate subsequent replanning.
\end{itemize}

\noindent
\textbf{Attack Evaluation.}
To quantify attack effectiveness, we employ a judge model $\mathcal{J}$. At each turn, the judge assigns a score $\mathcal{J}_t \in [1, 10]$ evaluating the extent to which $r_t$ fulfills the malicious intent of $q_{\text{harm}}$. The overall success score for the session is defined as the maximum score achieved: $J = \max_{t} \mathcal{J}(q_{\text{harm}}, r_t)$.

\subsection{Knowledge Accumulation}
\label{subsec:knowledge_engine}

To address the challenge of autonomous strategy discovery and eliminate the dependence on human-defined priors, We construct the knowledge repository $\mathbb{S}$ via a three-stage self-supervised pipeline: exploration, distillation, and integration. This process transforms a dataset of harmful queries $\mathcal{D}_{\text{aux}}$ into a compact set of diverse adversarial patterns without human supervision.

\noindent
\textbf{Self-Reflective Exploration.}
We employ a safety-aligned sandbox model as a proxy environment to bootstrap discovery. For a given query $q_{\text{aux}}$, the planner generates an initial exploratory plan $\mathcal{P}_0$ and executes it against the sandbox model. When an attack fails such that the judge score $J$ falls below the threshold $\tau$, the planner analyzes the interaction history $H_{\text{fail}}$ to diagnose why the response fell short and generates a revised plan $\mathcal{P}_{i+1}$: 

\begin{equation}
    \mathcal{P}_{i+1} \leftarrow \text{Planner}(\mathcal{T}_{\text{reflect}}(q_{\text{aux}}, \mathcal{P}_i, H_{\text{fail}}))
\end{equation}

This reflection enables the planner to diagnose the specific cause of failure and generate a targeted correction. This process ensures the generation of high-quality successful trajectories, thereby indirectly enhancing the effectiveness of the strategies subsequently extracted by the distiller.

\noindent
\textbf{Knowledge Distillation.} Once a successful jailbreak trajectory is achieved, we employ a distiller agent to decouple the contexual details from the underlying, transferable adversarial logic. Given the successful plan $\mathcal{P}_{\text{succ}}$ and the interaction history $H_{\text{succ}}$, the distiller identifies and extracts a set of atomic text-based attack strategies $\mathbb{S}_{\text{new}} = \{s_1, s_2, \dots, s_m\}$: 

\begin{equation}
    \mathbb{S}_{\text{new}} \leftarrow \text{Distiller}(\mathcal{T}_{\text{distill}}(q_{\text{aux}}, \mathcal{P}_{\text{succ}}, H_{\text{succ}}))
\end{equation}

These distilled strategies are then embedded and stored in the knowledge repository.

\noindent
\textbf{Knowledge Integration.} 
Upon completing the exploration of the auxiliary dataset, we collect a large volume of strategies with significant semantic redundancy. To transform this raw collection into a compact knowledge repository, we implement a process to progressively condense the strategy space.

Let $\mathbb{S}^{(t)}$ denote the set of strategies at iteration $t$:

\begin{itemize}
    \item \textbf{Clustering: }
    We compute the embedding vectors for all strategies in $\mathbb{S}^{(t)}$ and partition them into $K$ clusters $\{C_1, C_2, \dots, C_K\}$ via cosine similarity.
    
    \item \textbf{Consolidation: }
    For each cluster $C_k$, the distiller agent analyzes the enclosed strategies to identify both redundant duplicates and unique functional variants, then synthesizes a refined subset of strategies $\mathbb{S}'_k$:
    \begin{equation}
        \mathbb{S}'_k \leftarrow \text{Distiller}(\mathcal{T}_{\text{merge}}(C_k))
    \end{equation}
\end{itemize}

The repository is updated as the union of these refined subsets: $\mathbb{S}^{(t+1)} = \bigcup_{k=1}^{K} \mathbb{S}'_k$. The process repeats until the strategy count stabilizes.

\subsection{Knowledge-Driven Fuzzing}
\label{subsec:strategy_fuzzing}

Following the accumulation of the strategy repository, Mastermind transitions to the active deployment phase. We employ a grey-box fuzzing approach to identify optimal strategy combinations that trigger jailbreaks.

\noindent
\textbf{Fuzzing Formulation.}
We define the fuzzing campaign as an optimization task within the discrete strategy space. Let $\mathbb{S} = \{s_1, \dots, s_N\}$ denote the global repository of distilled strategies. We represent a specific test case as a binary vector $x \in \{0, 1\}^N$, where $x_k=1$ indicates the activation of strategy $s_k$. The objective is to identify the configuration $x^*$ that maximizes the vulnerability score determined by the execution oracle $\mathcal{F}$:

\begin{equation}
    x^* = \operatorname*{arg\,max}_{x \in \{0, 1\}^N} \mathcal{F}(x; q_{\text{target}})
\end{equation}

Here, the oracle $\mathcal{F}(x)$ instantiates the strategy vector $x$ into a concrete attack plan, executes the multi-turn interaction, and returns the maximum severity score assigned by the Judge.

\noindent
\textbf{Seed Scheduling.}
To accelerate the discovery process, we employ semantic seed scheduling rather than random initialization. We retrieve historical queries that are semantically similar to the current target $q_{\text{target}}$ using sentence embeddings. The successful configurations associated with the top-$K$ nearest neighbors form the initial population $\mathcal{P}_0$, ensuring the search begins in a region of high potential lethality.

\noindent
\textbf{Fuzzing Engine.}
To explore the input space and generate novel test cases, we implement an iterative evolution loop consisting of three sequential operators:

\begin{itemize}
    \item \textbf{Selection.} 
    We prioritize test cases that have demonstrated higher vulnerability scores. Through tournament selection, we identify high-performing parent configurations to drive the evolution of the next generation.
    
    \item \textbf{Crossover.} 
    This operator synthesizes new attack logic by merging patterns from distinct parents. Given parent vectors $x_a$ and $x_b$, we generate a hybrid offspring $\tilde{x}$ where each bit is inherited from either parent with equal probability:
    \begin{equation}
        \tilde{x}_k = \begin{cases}
            (x_a)_k & \text{with probability } 0.5 \\
            (x_b)_k & \text{with probability } 0.5
        \end{cases}
    \end{equation}
    Functionally, this facilitates the composition of partial solutions. It tests whether the synergy between effective substructures from different ancestors can yield a more coherent and potent attack plan.

    \item \textbf{Mutation.} 
    To ensure input diversity, we apply stochastic perturbation to the offspring. Each strategy bit in $\tilde{x}$ is flipped with a probability $p_m$:
    \begin{equation}
        x'_k = \begin{cases}
            1 - \tilde{x}_k & \text{with probability } p_m \\
            \tilde{x}_k & \text{otherwise}
        \end{cases}
    \end{equation}
    By stochastically injecting new strategies or pruning existing ones, it generates diverse test case variants to evaluate how localized adjustments to the attack surface can exploit potential gaps in the model’s safety guardrails.

\end{itemize}

\section{Experiments}
\subsection{Experiment Setting}

\noindent
\textbf{Datasets.} 
We evaluate the effectiveness of Mastermind using two distinct benchmarks: HarmBench~\cite{mazeika2024harmbench} and StrongReject~\cite{souly2024strongreject}. Specifically, we utilize the standard subset from HarmBench, which comprises 200 distinct harmful behavior instances covering diverse categories including cybercrime, chemical threats, hate speech, and misinformation. We further employ the custom small dataset from StrongReject, comprising 60 manually crafted instances that feature more natural and realistic behavioral details. In our experiments, we utilize the full datasets for the Main Results(Sec \ref{subsec:main_results}) and Attack Under Defense(Sec \ref{subsec:defense}), while employing a stratified random subset of 50 instances from HarmBench for the Ablation Study(Sec \ref{subsec:ablation_study}). 

To separate the testing data from the data used for knowledge accumulation phase, we constructed a separate auxiliary dataset containing 200 harmful queries generated via a commercial LLM, spanning similar semantic categories but containing no overlapping instances with the test sets.

\noindent
\textbf{Target Models.} We select a comprehensive suite of LLMs as victim models, ranging from open-source weights to proprietary commercial APIs. We categorize these targets into two groups: 
\begin{itemize} 
    \item \textbf{Standard Models:} This category includes widely deployed models such as the Llama 3 series (7B, 70B), Qwen 2.5 series (7B, 14B, 72B), DeepSeek V3, GPT-4o, and GPT-4.1. 
    \item \textbf{Reasoning Models:} Recognizing that chain-of-thought reasoning can potentially enhance safety robustness, we include models explicitly designed for reasoning, including DeepSeek R1, o3-mini, o4-mini, Gemini 2.5 Flash, Gemini 2.5 Pro, Claude 3.7 Sonnet, and GPT-5.
\end{itemize}

\noindent 
\textbf{Evaluation Metrics.} Following existing practices in recent jailbreak research, we employ the \textit{LLM-as-Judge} framework, utilizing GPT-4o to evaluate target models' responses. We report two complementary metrics:

\begin{itemize} \item \textbf{Attack Success Rate (ASR):} This is our primary metric. It functions as a binary judgment determining whether the response provides actionable information that satisfies the malicious intent without refusal.

\item \textbf{Harmfulness Rating (HR):} To capture the granularity of severity, we adopt a graded scoring (scale 0-5) following previous works \cite{kuo2025h}. This rating reflects the potential harm severity and toxicity of the generated content. \end{itemize}

It is worth noting that ASR and HR are distinct metrics with no inherent dependency. While they exhibit a positive correlation in some cases as successful jailbreaks typically yield harmful content, they capture different dimensions of model behavior and may diverge in specific scenarios.

\noindent
\textbf{Implementation Details.} We set the interaction bugdets $T_{\max} = 10$ turns. We instantiate the planner using DeepSeek V3 and the executor using Gemma 3 27B. During the knowledge accumulation phase, we utilize Llama 3.3 70B Instruct as the sandbox target for knowledge accumulation. All interactions are conducted with a temperature of 0 for the attacker agents and victim models to avoid the impact of randomness on experimental results.

\renewcommand{\arraystretch}{1} 
\begin{table*}[!t]
\caption{Performance Comparison of Multi-turn Jailbreaking Methods Across Different Target LLMs on the HarmBench Dataset.}
\label{tab:model_comparison}
\small
\centering
\setlength{\tabcolsep}{6pt} 
\begin{tabular}{lc cccc c} 
\toprule[1.5pt]
\textbf{Target Model} & \textbf{Metric} & \textbf{Crescendo} & \textbf{ActorAttack} & \textbf{X-Teaming} & \textbf{Siren} & \textbf{Mastermind} \\ 
\midrule[1.0pt]
\multirow{2}{*}{Llama 3.1 8B Instruct} 
 & ASR & 45\% & 30\% & 64\% & 43\% & \textbf{96\%} \\
 & HR & 2.75 & 2.81 & 3.12 & 2.19 & 3.42 \\
\midrule[0.4pt]
\multirow{2}{*}{Llama 3.3 70B Instruct} 
 & ASR & 23\% & 35\% & 79\% & 28\% & \textbf{92\%} \\
 & HR & 2.66 & 2.69 & 3.24 & 1.47 & 3.41 \\
\midrule[0.4pt]
\multirow{2}{*}{Qwen 2.5 7B Instruct} 
 & ASR & 65\% & 49\% & 79\% & 70\% & \textbf{89\%} \\
 & HR & 3.30 & 3.26 & 3.71 & 3.56 & 3.95 \\
\midrule[0.4pt]
\multirow{2}{*}{Qwen 2.5 14B Instruct} 
 & ASR & 30\% & 37\% & 78\% & 53\% & \textbf{92\%} \\
 & HR & 1.91 & 2.49 & 3.77 & 2.70 & 4.02 \\
\midrule[0.4pt]
\multirow{2}{*}{Qwen 2.5 72B Instruct} 
 & ASR & 40\% & 38\% & 74\% & 74\% & \textbf{95\%} \\
 & HR & 2.10 & 3.30 & 3.60 & 3.68 & 4.10 \\
\midrule[0.4pt]
\multirow{2}{*}{Deepseek V3} 
 & ASR & 28\% & 29\% & 79\% & 70\% & \textbf{94\%} \\ 
 & HR & 1.55 & 1.54 & 3.81 & 3.79 & 4.12 \\
\midrule[0.4pt]
\multirow{2}{*}{GPT-4o} 
 & ASR & 44\% & 28\% & 81\% & 56\% & \textbf{93\%} \\
 & HR & 2.74 & 2.56 & 3.78 & 3.27 & 3.97 \\
\midrule[0.4pt]
\multirow{2}{*}{GPT-4.1} 
 & ASR & 61\% & 30\% & 74\% & 71\% & \textbf{93\%} \\
 & HR & 3.16 & 2.81 & 3.61 & 3.41 & 3.94 \\
\midrule[1.0pt]
\multirow{2}{*}{Deepseek R1} 
 & ASR & 25\% & 22\% & 70\% & 67\% & \textbf{89\%} \\
 & HR & 1.60 & 1.31 & 3.74 & 3.60 & 4.04 \\
\midrule[0.4pt]
\multirow{2}{*}{o3 Mini} 
 & ASR & 24\% & 27\% & 74\% & 56\% & \textbf{90\%} \\
 & HR & 1.44 & 2.65 & 3.64 & 2.70 & 3.86 \\
\midrule[0.4pt]
\multirow{2}{*}{o4 Mini} 
 & ASR & 25\% & 40\% & 60\% & 56\% & \textbf{78\%} \\
 & HR & 2.76 & 2.94 & 3.47 & 2.99 & 3.59 \\
\midrule[0.4pt]
\multirow{2}{*}{Gemini 2.5 Flash} 
 & ASR & 43\% & 33\% & 70\% & 69\% & \textbf{91\%} \\
 & HR & 2.32 & 2.92 & 3.90 & 3.53 & 3.99 \\
\midrule[0.4pt]
\multirow{2}{*}{Gemini 2.5 Pro} 
 & ASR & 36\% & 38\% & 79\% & 73\% & \textbf{90\%} \\
 & HR & 1.98 & 3.31 & 3.99 & 3.84 & 4.08 \\
\midrule[0.4pt]
\multirow{2}{*}{Claude 3.7 Sonnet} 
 & ASR & 41\% & 14\% & 52\% & 21\% & \textbf{67\%} \\
 & HR & 2.95 & 1.58 & 3.26 & 1.28 & 3.42 \\
\midrule[0.4pt] 
 \multirow{2}{*}{GPT-5} 
 & ASR & 28\% & 16\% & 44\% & 18\% & \textbf{60\%} \\
 & HR & 2.48 & 1.24 & 3.43 & 1.37 & 3.71 \\
\midrule[1.0pt]
\multirow{2}{*}{\textbf{Average}} 
 & ASR & 37\% & 31\% & 70\% & 55\% & \textbf{87\%} \\
 & HR & 2.38 & 2.49 & 3.60 & 2.89 & 3.84 \\
\bottomrule[1.5pt]
\end{tabular} 
\end{table*}

\subsection{Main Results}
\label{subsec:main_results}

We present a comprehensive evaluation of Mastermind against four state-of-the-art multi-turn jailbreak baselines across 15 distinct LLMs. By analyzing performance on two distinct benchmarks, we demonstrate both the broad efficacy and the generalization capability of our framework.

\noindent
\textbf{Performance on HarmBench.}
We first evaluate Mastermind on the standard HarmBench dataset, where it demonstrates superior efficacy with an average Attack Success Rate (ASR) of 87\% across all tested models, significantly surpassing the strongest baseline, X-Teaming, which averages 70\%. 

Mastermind exhibits remarkable robustness across diverse model families. On standard open-weight and commercial models, such as DeepSeek V3 and GPT-4o, our method achieves ASRs of 94\% and 93\% respectively, highlighting a significant performance disparity compared to baselines like ActorAttack at 28\% and Crescendo at 44\%. Furthermore, Mastermind proves highly resilient against reasoning models, achieving 89\% on DeepSeek R1 and 90\% on o3-mini, whereas competing methods struggle to navigate the complex internal thought processes of these targets. Even against the most resilient next-generation targets, Mastermind remains the most effective framework. It is the sole method to sustain an ASR of 60\% against GPT-5, demonstrating superior adaptability to cutting-edge safety alignment compared to X-Teaming at 44\% and Siren at only 18\%.

\noindent
\textbf{Performance on StrongReject.} To validate the robustness of our attack across different data distributions, we further evaluated Mastermind on the StrongReject dataset. Consistent with the HarmBench findings, Mastermind achieves the highest ASR on every target model, culminating in an average ASR of 91\%. This consistently outperforms X-Teaming and Crescendo, which achieve 82\% and 79\% respectively. Notably, on the Qwen 2.5 72B model, Mastermind maintains a 93\% success rate while Siren's performance drops significantly to 67\%. These results confirm that Mastermind's effectiveness is not dataset-dependent and holds up under stricter evaluation criteria.

\begin{table*}[!t]
\caption{Performance Comparison of Multi-turn Jailbreaking Methods Across Different Target LLMs on the StrongReject Dataset.}
\label{tab:evaluation_summary_expanded}
\small
\centering
\setlength{\tabcolsep}{6pt} 
\begin{tabular}{lc cccc c} 
\toprule[1.5pt]
\textbf{Target Model} & \textbf{Metric} & \textbf{Crescendo} & \textbf{ActorAttack} & \textbf{X-Teaming} & \textbf{Siren} & \textbf{Mastermind} \\ 
\midrule[1.0pt]
\multirow{2}{*}{Llama 3.3 70B Instruct} 
 & ASR & 82\% & 68\% & 85\% & 34\% & \textbf{90\%} \\
 & HR & 3.01 & 3.08 & 3.30 & 1.97 & 3.65 \\
\midrule[0.4pt]
\multirow{2}{*}{Qwen 2.5 72B Instruct} 
 & ASR & 83\% & 68\% & 85\% & 67\% & \textbf{93\%} \\
 & HR & 3.20 & 3.18 & 3.47 & 3.42 & 3.80 \\
\midrule[0.4pt]
\multirow{2}{*}{Gemini 2.5 Flash} 
 & ASR & 83\% & 75\% & 80\% & 63\% & \textbf{88\%} \\
 & HR & 3.58 & 3.60 & 3.73 & 3.73 & 3.93 \\
\midrule[0.4pt]
\multirow{2}{*}{GPT-4o} 
 & ASR & 80\% & 67\% & 88\% & 72\% & \textbf{93\%} \\
 & HR & 3.08 & 3.32 & 3.77 & 3.25 & 3.85 \\
\midrule[0.4pt]
\multirow{2}{*}{o3 Mini} 
 & ASR & 68\% & 63\% & 72\% & 70\% & \textbf{90\%} \\
 & HR & 2.92 & 3.12 & 3.68 & 3.25 & 3.72 \\
\midrule[1.0pt]
\multirow{2}{*}{\textbf{Average}} 
 & ASR & 79\% & 68\% & 82\% & 61\% & \textbf{91\%} \\
 & HR & 3.16 & 3.26 & 3.59 & 3.12 & 3.79 \\
\bottomrule[1.5pt]
\end{tabular} 
\end{table*} 

\noindent
\textbf{Transferability and Generalization.}
A critical insight from our experiments concerns the transferability of the distilled strategies. During the knowledge accumulation phase, we utilized Llama 3.3 70B Instruct exclusively as the sandbox model to discover and distill attack patterns. As expected, Mastermind achieves exceptional performance on this source architecture, recording a 92\% ASR. 

However, this high performance is not limited to the target model. Mastermind achieves comparable, and in several cases superior, performance on completely disparate architectures, attaining 95\% on Qwen 2.5 72B, 94\% on DeepSeek V3, and 93\% on GPT-4o. This phenomenon indicates that our distiller agent successfully decouples adversarial logic from model-specific artifacts. Instead of overfitting to the idiosyncrasies of the Llama 3.3 sandbox, Mastermind extracts generalized semantic vulnerabilities that remain effective across different model families and alignment methodologies. This confirms that the strategies stored in our knowledge repository possess strong transferability, allowing the framework to compromise unknown black-box models effectively.

\begin{figure}[!h]
    \centering
    \includegraphics[width=\linewidth]{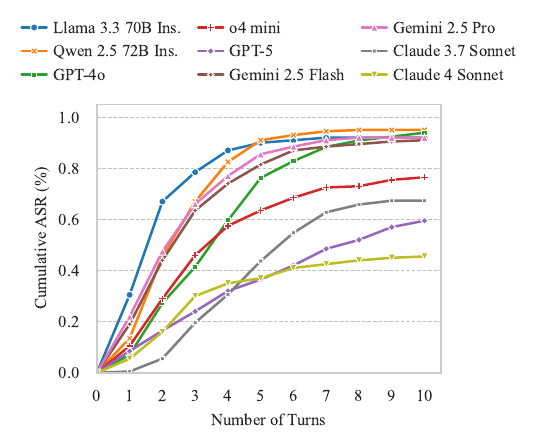}
    \caption{Cumulative ASR as a function of interaction turns across various target LLMs. The x-axis represents the progression of the dialogue from turn 0 to 10 , while the y-axis indicates the ASR by that stage.}
    \label{fig:asr_plot}
\end{figure}

\noindent
\textbf{Severity and Progression.}
Finally, beyond success rates, the qualitative impact of the attacks is substantial. Mastermind achieves the highest average Harmfulness Rating (HR) of 3.84 on HarmBench and 3.79 on StrongReject. This indicates that the successful attacks yield highly detailed and actionable instructions rather than trivial bypasses. As illustrated in Figure \ref{fig:asr_plot}, while Mastermind saturates standard models within the first few turns, it utilizes the full turn budget to gradually erode the defenses of resilient targets like GPT-5, underscoring the necessity of the multi-turn strategy in overcoming advanced safety guardrails.

\subsection{Ablation Study}
\label{subsec:ablation_study}
\textbf{Contribution of Architectural Components.} 
To strictly isolate the contributions of Mastermind's hierarchical architecture and knowledge mechanism, we conducted a component-wise ablation study on GPT-5, our most resilient target. We established a baseline configuration consisting solely of the executor, functioning as a standard context-aware generator without strategic oversight.

As presented in Table \ref{tab:ablation_study}, the baseline struggles to breach the advanced safety alignment of GPT-5, achieving only a 30\% ASR. The introduction of the planner yields a significant performance boost, raising the ASR to 46\%. This improvement validates the necessity of decoupling high-level strategy from low-level execution; by charting a coherent, step-by-step roadmap, the planner prevents the objective drift that plagues the baseline.

The integration of the controller enables the self-reflective mechanism (Phase 1). By analyzing responses and iteratively refining the plan in real-time—mirroring the exploration process used during knowledge accumulation—the system raises the ASR to 56\%. This demonstrates the value of dynamic error correction. Finally, the full framework (Phase 2) augments this system with the retrieval of accumulated knowledge. By initializing the planner with proven, distilled strategies rather than starting from scratch, Mastermind achieves a peak ASR of 66\% and a high HR of 3.78. This confirms that while reflection provides resilience, the knowledge repository provides the critical strategic priors.

\noindent
\textbf{Impact of the scale of knowledge repository.} To investigate the contribution of the accumulated adversarial knowledge to the attack effectiveness, we conducted an ablation study by varying the size of the knowledge repository. We randomly sampled subsets of the distilled strategies at ratios of 0\%, 20\%, 40\%, 60\%, 80\%, and 100\%, and evaluated Mastermind's performance on the benchmark.

\begin{figure}[!htbp]
    \centering
    \includegraphics[width=\linewidth]{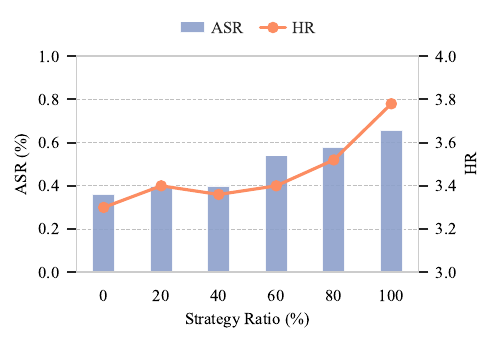}
    \caption{Performance under varying knowledge repository sizes. The blue bars represent the ASR on the left axis, and the orange line tracks the HR on the right axis, evaluated across strategy retention ratios ranging from 0\% to 100\%.}
    \label{fig:asr_vs_strategies}
\end{figure}

\begin{table}[!h]
\caption{Component ablation on GPT-5, verifying the gains from hierarchical planning, self-reflective exploration (Phase 1), and knowledge repository (Phase 2).}
\label{tab:ablation_study}
\small
\begin{tabular*}{\columnwidth}{@{\extracolsep{\fill}}l c r}
\toprule[1.5pt]
\textbf{Components} & \textbf{ASR} & \textbf{HR} \\
\midrule[1.0pt] 
Base                  & 30\% & 3.40 \\
+ Planner             & 46\% & 3.56 \\
+ Planner + Controller(Phase 1) & 56\% & 3.34 \\
+ Planner + Controller(Phase 1 + 2) & 66\% & 3.78 \\
\bottomrule[1.5pt]
\end{tabular*}
\end{table}
As illustrated in Figure \ref{fig:asr_vs_strategies}, the scale of the strategy repository shows a direct positive correlation with attack performance, validating the critical importance of our accumulated knowledge. Specifically, the ASR exhibits a consistent upward trend as the strategy availability increases. Without any historical strategies, the baseline ASR is approximately 36\%. In contrast, utilizing the full repository nearly doubles the success rate to 66\%. This substantial performance gap highlights that the distilled strategies are fundamental to the system's ability to compromise the target model effectively. The average HR follows a similar upward trajectory, rising from approximately 3.3 to 3.8. This indicates that a richer strategy pool does not only improve the frequency of successful attacks but also enhances the severity of the generated responses.

\begin{figure*}[!htbp]\centering\includegraphics[width=\linewidth]{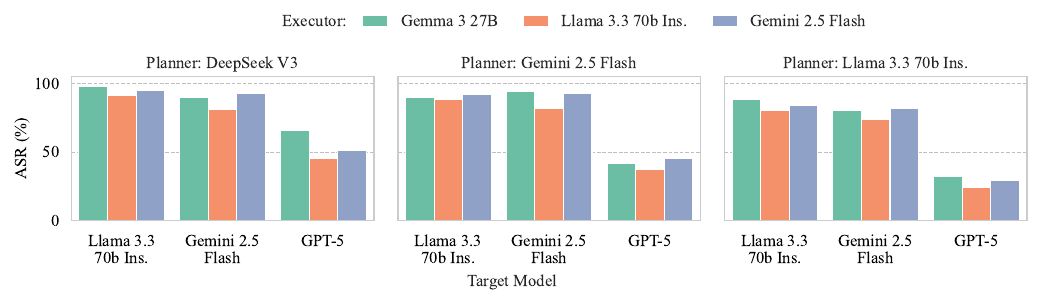}\caption{Impact of planner and executor model selection on ASR across target victim models.}\label{fig:ablation_on_models}\end{figure*}

These results strictly demonstrate that the knowledge repository is a key driver of Mastermind's performance. The high correlation confirms that the effectiveness of the fuzzing process relies heavily on the quality and quantity of the adversarial strategies distilled during the knowledge accumulation phase.

\noindent
\textbf{Impact of model selection for different agents.} We investigate the modularity of our framework by varying the LLMs instantiated as the planner and executor. Figure \ref{fig:ablation_on_models} illustrates the attack success rates across three distinct victim models.

First, the planner's reasoning capability is the primary determinant of success. DeepSeek V3 consistently outperforms other planners, especially when targeting the resilient GPT-5. It achieves approximately 65\% ASR in this scenario, compared to only around 30\% when Llama 3.3 70B is used as the planner. This indicates that the ability to decompose complex adversarial goals is more critical than the generative capacity required for execution.

Second, executor performance reflects an alignment bias with the knowledge source. Gemma 3 27B, which was used as the sandbox executor during knowledge accumulation, achieves the highest ASRs. For instance, it reaches 98\% against the Llama 3.3 70B target. However, the accumulated strategies demonstrates strong transferability. Even when using agents that were not involved in the discovery phase—such as employing Gemini 2.5 Flash as the planner or Llama 3.3 70B as the executor—the ASRs consistently exceed 75\% against standard targets. This confirms that the distilled strategies capture generalized adversarial logic rather than overfitting to specific model patterns.

\begin{table*}[!t]
\centering
\small
\caption{Performance with Different Defense Mechanisms Applied.}
\label{tab:defenses}
\setlength{\tabcolsep}{6pt} 
\renewcommand{\arraystretch}{1.0} 
\begin{tabular}{lc cccc}
\toprule[1.5pt]
\textbf{Target Model} & \textbf{Metric} & \textbf{W/O Defense} & \textbf{Self-Reminder} & \textbf{SmoothLLM} & \textbf{Llama Guard} \\
\midrule[1.0pt]
\multirow{2}{*}{Llama 3.3 70B Instruct} 
 & ASR & \textbf{92\%} & 77\% & 88\% & 86\% \\
 & HR & 3.41 & 2.96 & 3.42 & 3.44 \\
\midrule[0.4pt]
\multirow{2}{*}{Qwen 2.5 72B Instruct} 
 & ASR & \textbf{95\%} & 87\% & 91\% & 87\% \\
 & HR & 4.10 & 3.20 & 3.87 & 3.77 \\
\midrule[0.4pt]
\multirow{2}{*}{GPT-4o} 
 & ASR & \textbf{93\%} & 72\% & 85\% & 92\% \\
 & HR & 3.97 & 2.95 & 3.46 & 3.37 \\
\midrule[0.4pt]
\multirow{2}{*}{o3 Mini} 
 & ASR & \textbf{90\%} & 87\% & 84\% & 87\% \\
 & HR & 3.86 & 3.60 & 3.62 & 3.83 \\
\midrule[0.4pt]
\multirow{2}{*}{Gemini 2.5 Flash} 
 & ASR & \textbf{91\%} & 79\% & 79\% & 81\% \\
 & HR & 3.99 & 3.48 & 3.84 & 3.90 \\
\midrule[1.0pt]
\multirow{2}{*}{\textbf{Average}} 
 & ASR & \textbf{92\%} & 80\% & 85\% & 87\% \\
 & HR & 3.87 & 3.24 & 3.64 & 3.66 \\
\bottomrule[1.5pt]
\end{tabular}
\end{table*}

\subsection{Resilience against Defenses}
\label{subsec:defense}
To evaluate the robustness of Mastermind against existing defense mechanisms, we conduct experiments using three defense methods across five target models. Table \ref{tab:defenses} presents the performance comparison between our method with and without defense mechanisms deployed.

Existing representative defense mechanisms were primarily designed for single-turn interactions, requiring adaptation to protect against multi-turn attacks. We select and adapt three representative defenses:

\begin{itemize}
    \item \textbf{Self-Reminder:} Originally designed to wrap user instructions with safety prompts in single-turn scenarios, this defense incorporates explicit safety instructions into the system prompt. For multi-turn adaptation, we configure the safety prefix as the system prompt at conversation initialization and append the safety suffix to the user query in the last turn.
    
    \item \textbf{SmoothLLM:} This perturbation-based defense applies character-level modifications to detect jailbreak attempts through response consistency analysis. In our multi-turn setting, we apply perturbations specifically to the most recent user query before each interaction between the executor and the target model.
    
    \item \textbf{Llama Guard:} We employ Llama Guard 4, which features native conversation-level analysis capabilities. Before the executor transmits each message, Llama Guard 4 evaluates the entire conversation history for potential policy violations. If harmful content is detected, the interaction is treated as refusal.
\end{itemize}

The results indicate that while specific defenses can attenuate the attack performance, Mastermind retains a high level of effectiveness with the average ASR remaining above 80\%. Harmfulness rating reveals that our method not only sustains high success rates but also preserves the severity and detail of the generated content against active defenses.

Among the evaluated defenses, Self-Reminder proves to be the most effective countermeasure. It reduces the average ASR from 92\% to 80\% and causes the most significant drop in average HR from 3.87 to 3.24. This decline in HR is particularly notable in GPT-4o where the score falls to 2.95. This trend suggests that the explicit safety reminders in the system prompt induce a constraint on output severity, forcing the model to generate more guarded responses. However, an average HR of 3.24 indicates that the successful attacks still yield actionable harmful information.

SmoothLLM demonstrates limited efficacy in mitigating attack severity. While it marginally lowers the average ASR to 85\%, the average HR remains robust at 3.64. Specifically, models like Qwen 2.5 72B maintain a high severity score of 3.87 under this defense. This result highlights that the semantic core of the adversarial strategy remains intact even when input perturbations occur, ensuring that successful jailbreaks result in detailed outputs without significant degradation in quality.

The external filter Llama Guard yields an average ASR of 87\% and an average HR of 3.66 which indicates it is the least effective defense in this setting. The data demonstrates that this mechanism generally fails to significantly curtail either the success rate or the severity of the generated responses. For instance, the success rate for GPT-4o remains high at 94\% and the harmfulness ratings for models such as Llama 3.3 70B and o3 Mini remain comparable to the undefended baseline. This suggests that Mastermind can effectively bypass external classifiers that monitor conversation history.

Overall, these results suggest that current defense paradigms are insufficient to fully mitigate the threat. Mastermind demonstrates resilience by adapting its strategies to circumvent defenses while maintaining a high degree of content severity. This proves that the generated attacks remain practically dangerous even under defensive scrutiny.

\section{Conclusion}
In this paper, we introduced Mastermind, a knowledge-driven framework that autonomously uncovers vulnerabilities in Large Language Models through hierarchical multi-turn interactions. By decoupling strategic planning from tactical execution and employing a self-evolving strategy-space fuzzing mechanism, Mastermind effectively overcomes the limitations of existing rigid and incoherent attack methods. Comprehensive evaluations demonstrate its superior efficacy and transferability against state-of-the-art models, including GPT-5 and Claude 3.7 Sonnet, even under advanced defense mechanisms. Ultimately, this work highlights significant gaps in current safety alignment and establishes a robust, scalable paradigm for automated red-teaming to enhance future model resilience.

\cleardoublepage
\bibliographystyle{plain}
\bibliography{ref}

\end{document}